**Superconductivity at 48 K of heavily hydrogen-doped SmFeAsO epitaxial films grown by topotactic chemical reaction using CaH$_2$**

Jumpei Matsumoto,[1] Kota Hanzawa,[1] Masato Sasase,[2] Silvia Haindl,[3] Takayoshi Katase,[1] Hidenori Hiramatsu,[1,2,a)] and Hideo Hosono[2]

[1] *Laboratory for Materials and Structures, Institute of Innovative Research, Tokyo Institute of Technology, Mailbox R3-3, 4259 Nagatsuta-cho, Midori-ku, Yokohama, Kanagawa 226-8503, Japan*

[2] *Materials Research Center for Element Strategy, Tokyo Institute of Technology, Mailbox SE-1, 4259 Nagatsuta-cho, Midori-ku, Yokohama, Kanagawa 226-8503, Japan*

[3] *Tokyo Tech World Research Hub Initiative (WRHI), Institute of Innovative Research, Tokyo Institute of Technology, 4259 Nagatsuta-cho, Midori-ku, Yokohama, Kanagawa 226-8503, Japan*

High-critical-temperature ($T_c$) superconductivity at 48 K is reported for hydrogen-doped SmFeAsO epitaxial films on MgO single-crystal substrates. The key processes are pulsed laser deposition to grow undoped SmFeAsO epitaxial films and subsequent topotactic chemical reaction using CaH$_2$ powders under evacuated silica-glass ampule atmosphere. Based on this post-deposition thermal annealing treatment that we have developed, a maximum hydrogen concentration $x = \sim0.35$ was realized in SmFeAs(O$_{1-x}$H$_x$). Disordered hydrogen-substitution at O sites is experimentally confirmed directly by atomic-scale microstructural observations. Magnetization measurement validates the bulk nature of the high-$T_c$ superconductivity in the films. This method will become an effective and general method to fabricate various high-quality oxyhydride epitaxial films.

---

[a)] Author to whom correspondence should be addressed. Electronic mail: h-hirama@mces.titech.ac.jp

I. INTRODUCTION

In 2008, Kamihara *et al.* reported that approximately 10% F doping at the O site of LaFeAsO induces superconductivity with a critical temperature ($T_c$) = 26 K [1], which was 2 years after their report on superconductivity at $T_c$ = 4 K in LaFePO [2]. Both compounds crystallize in the ZrCuSiAs-type structure and are called 1111-type Fe-based superconductors. They are isomorphic to wide-gap *p*-type layered oxychalcogenide semiconductors ReCuOCh (Re = rare earth, Ch = chalcogen) [3, 4]. Usually, Fe with its large magnetic moment is believed to be harmful to the emergence of superconductivity and similar assumptions were also applied to other magnetic 3*d* transition metals such as Ni and Co. Thus, the above discovery of high-$T_c$ superconductivity at 26 K in F-doped LaFeAsO [1] was accepted with surprise in the condensed matter community and globally rekindled extensive exploration studies of new superconductors [5]. These materials have introduced a new class of high-$T_c$ superconductors and they have characteristic features for future practical applications such as superconducting wires, tapes, and coated conductors for high-field magnets [6–9].

The maximum $T_c$ among the Fe-based superconductors is ~55 K, which is reported for F-doped 1111-type SmFeAsO [10], called Sm1111. The substitution limit of fluorine is ~20% with respect to its O site. However, Hanna *et al.* [11] reported that oxygen can also be replaced by hydrogen with a maximum $T_c$ in H-doped Sm1111 [SmFeAs(O$_{1-x}$H$_x$)] comparable to that of F-doped Sm1111. The most important characteristics of H doping is that the substitution limit extends to ~80% (i.e., $x$ = 0.8), leading to the discovery of a unique electronic phase diagram among 1111-type ReFeAsO [11–13]. It is considered that in both cases, H-doped as well as F-doped SmFeAsO, superconductivity originates from electron doping by substitution at the O site. Although small size crystals with maximum at ~400 μm were reported recently [14], it is very difficult to grow enough large-size single crystals of H-doped Sm1111, which restricts further detailed investigations on the superconducting mechanism and the electronic band structure by experimental methods such as angle-resolved photoemission spectroscopy.

We therefore propose another synthesis approach in order to make future studies on H-doped Sm1111 accessible. Here, we report on large size heteroepitaxial thin-film samples on single-crystalline substrates with a usual lateral size of 10 mm × 10 mm. Just after the report by Kamihara *et al.* in 2008 [1], extensive synthesis and research did not only start on bulk samples (polycrystalline and single crystals) but also thin film growth – especially epitaxial films – was stimulated worldwide [8, 9, 15, 16]. Pulsed laser deposition (PLD) turned out to be a compatible technique also for Fe-based superconductors and has been employed mainly because of the historical success in the growth of high-$T_c$ cuprate thin films and coated conductors. It is important to note here that until today epitaxial ReFeAsO thin films were mainly grown by PLD equipped with a Nd:YAG laser. A KrF excimer laser ($\lambda$ = 248 nm) was only employed in a room-temperature deposition for chemical transfer before a post-deposition thermal annealing of La1111 films [17, 18]. Hiramatsu *et al.* [19] employed a Nd:YAG laser as an excitation for the PLD process instead of an ultraviolet excimer laser, initially demonstrating *in situ* La1111 heteroepitaxial growth. Using the same Nd:YAG PLD system, Haindl *et al.* [20–23] recently realized *in situ* PLD growth of high-quality undoped and F-doped Sm1111 epitaxial films. We want to point out here, that at present the only known successful attempts of 1111-type iron oxypnictide thin film growth by *in situ* PLD were demonstrated using a Nd:YAG laser. The influence of the excitation laser wavelength in PLD of iron pnictides is still unclear for the 1111 system while the relevant research was investigated only for 122-type $BaFe_2As_2$ [24]. At present, besides PLD, molecular beam epitaxy (MBE) is a common technique for the *in situ* growth of 1111-type ReFeAsO films [25, 26], and metal-organic chemical vapor deposition has made progress in the fabrication of 1111-type ReFeAsO films [27].

Our studies on *in situ* PLD growth of ReFeAsO imply that the optimization of growth conditions is severe irrespective of use of a Nd:YAG laser for ablation [21, 22] and any attempts to achieve *in situ* PLD growth of H-doped Sm1111 epitaxial films failed so far. Even our initial efforts in a reactive PLD process demonstrated that hydrogen could not be doped into epitaxial Sm1111 films when $H_2$ gas and rf-generated H radical gas were introduced into the growth chamber during deposition experiments, which will be discussed later in Fig. S1 [28]. The essential key in this study is a topotactic chemical reaction between undoped Sm1111 epitaxial precursor films, grown by the Nd:YAG PLD, and $CaH_2$ powder using a unique post-deposition thermal annealing experimental setup.

II. EXPERIMENTAL

Figure 1(a) shows the growth process of undoped Sm1111 heteroepitaxial thin films. PLD using the second harmonics ($\lambda$ = 532 nm) of a Nd:YAG laser was employed to ablate undoped Sm1111 polycrystalline disks. The films were deposited on MgO single-crystalline substrates (size: 10 mm × 10 mm × 0.5 mm), heated at 900 ºC with an infrared semiconductor laser, under a high vacuum atmosphere (base pressure: $\leq 1\times10^{-6}$ Pa). The MgO substrate was thermally annealed at 1050 ºC in air before deposition. The thicknesses of all the undoped Sm1111 films were fixed to be ~80 nm. The Nd:YAG PLD-film-growth procedure was carried out according to the details reported in refs. [19–22]. These undoped Sm1111 epitaxial films grown by the Nd:YAG PLD are called precursor films (No. 1) and serve as starting point in this study.

Although we initially attempted H-doping of Sm1111 epitaxial films by *in situ* growth under $H_2$ gas and rf-generated H radical atmospheres as well as by post-deposition thermal annealing treatment at 800 ºC under $H_2$ atmosphere, no H-doped Sm1111 films could be obtained. To efficiently substitute Sm1111 epitaxial films with H, we therefore performed a topotactic chemical reaction using an active solid H-source, which is $CaH_2$ polycrystalline powder, through two different experimental approaches [Figs. 1(b) and 1(c)]. The topotactic chemical reaction we expected for both was

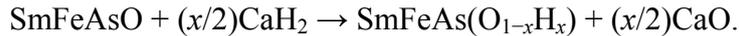
$$SmFeAsO + (x/2)CaH_2 \rightarrow SmFeAs(O_{1-x}H_x) + (x/2)CaO.$$

Figure 1(b) represents the first approach, which is based on our previous reports on heteroepitaxial growth of materials composed of high vapor pressure elements such as K [29, 30]. The precursor film (No. 1) was put at the bottom of an alumina vessel, which was preliminary paved by large amounts of $CaH_2$ powder to keep high H vapor pressure during subsequent thermal annealing process, and then capped with a Y-stabilized $ZrO_2$ (YSZ) single-crystal plate to protect the surface of the undoped Sm1111 film. More $CaH_2$ powder was put on the YSZ plate, and then put on an alumina top cover. Finally, the alumina vessel was thermally annealed in an evacuated silica-glass ampule at 500 ºC for 24 h. The resulting H-doped film is defined as sample 2. The second approach resulted in heavy H-doped films (No. 3) that were fabricated through two steps similar to those in No. 2 as shown in Fig. 1(c). Here, the precursor film (No. 1) was directly filled together with large amounts of $CaH_2$ powder (~1.5 g) in silica-glass tube and subsequently sealed under a vacuum atmosphere in silica-glass ampule, i.e., this approach does not employ a film surface protection with YSZ and efficiently promotes the topotactic reaction and diffusion of hydrogen into the film. The final thermal annealing process condition is carried out under the same conditions as that for No. 2. Note that the annealing temperature of 500 ºC is the optimum because lower (400 ºC) and higher (600 ºC) temperatures were not effective to induce superconductivity, which will be discussed later on.

The crystalline phases and structure parameters such as lattice parameters were characterized by x-ray diffraction (XRD) with monochromated Cu K$\alpha_1$ radiation. The crystallinity and crystallographic orientation of resulting Sm1111 films were evaluated by x-ray rocking-curve measurements at the Sm1111 003 diffraction for the out-of-plane direction and at the Sm1111 200 diffraction for the in-plane direction. Scanning transmission electron microscope (STEM) with high-angle annular dark field (HAADF) and annular bright-field (ABF) modes was employed for the observation of atomic-scale cross-sectional microstructure. H concentration [$x$ in SmFeAs(O$_{1-x}$H$_x$)] of the H-doped Sm1111 epitaxial films was quantitatively determined from depth profiles obtained by secondary ion mass spectroscopy (SIMS) using a polycrystalline SmFeAs(O$_{0.4}$H$_{0.6}$) bulk sample [13] as a standard sample. Temperature ($T$) dependence of the electrical resistivity ($\rho$) was measured using four-probe method with Au electrodes from 2 to 300 K under magnetic fields up to 9 T in a commercial Physical Properties Measurement System (PPMS). A constant current of 5 mA was used in the electrical transport measurements. Magnetization was characterized at $T$ = 4−300 K using zero-field cooling (ZFC) and field cooling (FC) at 10 Oe to confirm the bulk superconductivity in the H-doped Sm1111 films Nos. 2 and 3. ZFC and FC magnetizations were obtained with a Vibrating Sample Magnetometer, which is an option of PPMS.

III. RESULTS AND DISCUSSION

*III-1. Thin-film growth by Nd:YAG PLD and thermal annealing*

Before we employed the topotactic chemical reaction using CaH$_2$ powder as an active H-source for H doping, we preliminarily attempted H doping during reactive *in situ* PLD film-growth as well as by post-deposition thermal annealing under H$_2$ atmosphere. Figure S1 [28] summarizes the resultant XRD patterns and $\rho$−$T$ curves in each employed process. Figigures S1(a) and S1(c) [28] display H pressure ($P_{H2}$) dependence of obtained crystalline phases, where $H_2$ gas and rf-generated H radical gas were introduced into the vacuum chamber for PLD growth during the film growth experiments, respectively. In both processes, impurity phases such as SmAs segregated with the increase in $P_{H2}$. Furthermore, their $\rho-T$ behavior stayed insulator like [19, 20, 31 – 33] as shown in Figs. S1(b) for $H_2$-gas growth and S1(d) for growth with the rf-H radical source [28]. Then, we performed post-deposition thermal annealing under $H_2$ gas atmosphere for as-grown Sm1111 epitaxial thin films, which were deposited under high vacuum at 800 ºC. Figures S1(e) and S1(f) [28] are XRD patterns and $\rho-T$ curves before and after the annealing, respectively. Similar to both above attempts of a reactive *in situ* growth, this post-deposition annealing process could not induce a superconducting transition either. These results suggested that hydrogen was not substituted for oxygen probably due to the strong chemical bonding between O and Sm within the O–Sm layers of the layered crystal structure.

*III-2. Topotactic chemical reaction using $CaH_2$*

Thus we performed a post-deposition thermal annealing on undoped precursor films using the binary hydride $CaH_2$ as a more active solid H-source in order to achieve high partial pressures (i.e., chemical potential) of hydrogen during the chemical reaction that finally result in hydrogen substitution for oxygen of undoped Sm1111 epitaxial films preliminarily grown by PLD. Figure 2(a) shows out-of-plane XRD patterns of three Sm1111 films (wider $2\theta$ range patterns are shown in Fig. S2(a) [28]. The annealing temperature of 500 ºC is found to be optimum because annealing at lower (400 ºC) and at higher (600 ºC) temperatures were not effective to induce superconductivity. Figure S3 [28] shows a thermal desorption spectrum of the $CaH_2$ powder. Desorption of the hydrogen molecule from $CaH_2$ starts around 400 ºC and then exhibits a maximum value at ~620 ºC. Therefore, 400 ºC is too low for doping enough hydrogen in the films. At 600 ºC, the thermal desorption of hydrogen is more effective than at 500 ºC. However, the main Sm1111 crystalline phase of the precursor films was completely decomposed at 600 ºC (see Fig. S4 [28]). A similar decomposition around 600 ºC – 800 ºC was observed also in Fig. S1(e) as well as in La1111 epitaxial films grown by Nd:YAG PLD [19].

Undoped Sm1111, No. 1, exhibited strong *c*-axis orientation without any impurity phases for out-of-plane. This feature did not change in thermally annealed samples of Nos. 2 and 3. Figure 2(b) displays out-of-plane rocking curves of the 003 diffraction. The full widths at half maxima are all ~0.6º. Figures 2(c) and 2(d) show the in-plane XRD patterns (wider $2\theta\chi$ range patterns are shown in Fig. S2(b) [28]) and in-plane rocking curves, respectively. For Nos. 1 and 3, only diffraction peaks of the Sm1111 phase and the MgO substrate were observed [Fig. 2(c)], whereas for No. 2, an impurity phase (indicated by a filled circle) was detected. A possible candidate could be the 400 diffraction of $As_2O_5$. Because from the in-plane four-fold symmetry reflecting the tetragonal lattice of Sm1111 observed in Fig. 2(d), we concluded that the Sm1111 lattice maintained a heteroepitaxial relation with the MgO substrate, $[001]_{SmFeAsO} \parallel [001]_{MgO}$ for out-of-plane and $[100]_{SmFeAsO} \parallel [100]_{MgO}$ for in-plane, which remains unchanged under additional thermal annealing and the topotactic chemical reaction. From these XRD patterns, we estimated the *a*- and *c*-axis lattice parameters. For the undoped precursor film (No. 1), we would like to note that an *a*-axis lattice parameter of 3.923 Å irrespective of the larger in-plane lattice parameter of MgO substrate ($a$ = 4.213 Å), and a *c*-axis lattice parameter of 8.561 Å, clearly different to the lattice parameters of undoped polycrystalline bulk ($a$ = 3.9395 and $c$ = 8.4969 Å [11]), originating from lattice strain due to heteroepitaxial growth. In the case of H doping of Sm1111 polycrystalline bulks, the lattice parameters continuously decrease both along *a* and *c* axes with increasing H concentration *x* [11]. Accordingly, the lattice parameters of our epitaxial films also shrank in the order from Nos. 1 and 2 ($a$ = 3.912 and $c$ = 8.534 Å) to No. 3 ($a$ = 3.903 and $c$ = 8.488 Å). This implies indirectly that an increasing amount of H is incorporated into the Sm1111 epitaxial films with increasing sample number.

*III-3. Atomic-scale microstructure observation*

We then observed atomic-scale microstructure of the sample #3, in which the highest H concentration was expected based on the above lattice parameter analysis. Figure 3(a) shows a wide-view cross-sectional microstructure. Originating from its layered crystal structure, alternately stacking patterns along the *c* axis were clearly observed from interface to the film surface on this scale without any noticeable defects such as dislocations, validating its high quality. The inset of Fig. 3(a) shows a very sharp interface (heteroepitaxial growth starts within a few atomic layers from the interface), indicating that any chemical reaction between the film and the MgO does not occur during the *in situ* PLD growth and the post-deposition thermal annealing. The HAADF-STEM image [Fig. 3(b)] allows to atomically resolve the 1111-type layered structure. However, we could not observe a clear Z-contrast originating from H substitution at the O site in this image because the contrast of neighboring Sm, which bonds with O, was too bright compared to the O signals. It should be noted that in Fig. 3(c) we attained clear experimental evidence for hydrogen substitution at the O site by ABF-STEM. In this figure, originating from its bright-field characteristics, the darkest and largest ones are Sm positions, and some O atoms at O sites are indicated by yellow arrows. We found that brighter positions indicated by light blue arrows also exist at the same O sites. This observation indicates that doped H substitutes at the O site and it is completely disordered. This substitution is also confirmed in the intensity plot at the bottom of Fig. 3(c).

*III-4. H concentration in the films*

According to the above long-range (i.e., entirely averaged in XRD) and local atomic-scale structural analyses by STEM, H doping of Sm1111 epitaxial films was successfully introduced by the $CaH_2$ route. Hence, we then quantitatively analyzed H concentrations [$x$ in $SmFeAs(O_{1-x}H_x)$] in all the samples by SIMS. Figures 4(a)−4(c) show SIMS depth profiles for samples 1, 2, and 3, respectively. Figure 4(d) is a summary of these profiles in terms of hydrogen signal. Although the H signal for No. 1 was almost comparable to the detection limit for SIMS, clear enhancement of $x$ was detected in both films 2 and 3. The maximum $x$ in each sample was ~0.35 for No. 3 (averaged $x$ = 0.19 along the film thickness area) and ~0.12 for No. 2 (averaged $x$ = 0.08) around the film surface [see the inset of Fig. 4(d)]. This result validates that a topotactic chemical reaction occurs as we expected. We also tried to clarify the relevant chemical reactions between the as-grown Sm1111 film and $CaH_2$. Figure S5 [28] shows the powder XRD patterns of $CaH_2$ before and after annealing. We could not observe any diffraction peak originating from a crystalline CaO phase. This is due to the following reasons: First, the mass of the films is much smaller than that of the $CaH_2$ powder (~1.5 g), second, there is also the possibility that an amorphous phase formed, which could not be detected by XRD. We think that the most plausible reaction path for H doping is:

$$\text{SmFeAsO} + (x/2)\text{CaH}_2 \rightarrow \text{SmFeAs}(\text{O}_{1-x}\text{H}_x) + (x/2)\text{CaO}.$$

Additionally, as observed in Fig. 4, the oxygen content did not change in samples 1 and 2. In sample 3 the oxygen content is slightly reduced probably because chemical reaction with $\text{CaH}_2$. At present we consider this chemical reaction as reliable.

For sample 2, the hydrogen concentration is much lower than that of No. 3, indicating that the thermal diffusion of hydrogen was clearly slowed down by the use of a YSZ protection cap. Additionally, an inhomogeneous $x$ distribution along the film-thickness direction is also observed, most strongly for No. 3. Although a hydrogen diffusion gradient is observed, hydrogen diffused entirely from the surface to the film/substrate interface. Figure 5 shows the relationship between the determined H concentration, $x$, and the lattice parameters. Similar to the lattice parameter in H-doped Sm1111 bulk samples [11], both $a$- and $c$-axis lattice parameters of H-doped Sm1111 epitaxial thin films (Nos. 2 and 3) decreased with an increase in $x$, suggesting that the incorporated amount of H continuously substituted O and electron doping can be expected.

*III-5. Superconducting properties*

Next, we measured the electronic transport properties of samples 1 − 3. Figures 6(a) and 6(b) show $\rho-T$ curves in temperature ranging between 2−300 K and 35−54 K, respectively. The undoped Sm1111 epitaxial precursor film (No. 1) exhibited insulator-like behavior, which has been also observed in other 1111-type epitaxial thin films [19, 20, 31 − 33], in contrast to a metallic resistivity in undoped Sm1111 bulks [11]. According to its undoped characteristics, an anomaly in $\rho-T$ was observed as indicated by the vertical arrow ($T_N$), which is probably due to an antiferromagnetic transition similar to that found in bulk samples [11]. In the H-doped films (Nos. 2 and 3) a clear metallic resistivity was observed for temperatures above 50 K. We performed a power-law fitting for the normal states using an equation $\rho = \alpha + \beta T^n$, where $\alpha$ and $\beta$ are constants. The exponent $n$ values in the normal-conducting state were 1.1 for No. 2 and 0.6 for No. 3, which are roughly consistent with those found in bulk samples (1.15 at $x = 0.13$ and 0.88 at $x = 0.34$) [11].

For temperatures lower than 50 K, sharp drops of $\rho$ at 45.2 K for No. 2 and 47.9 K for No. 3 indicated a superconducting transition. Although zero resistance was not detected down to 2 K in No. 2 owing to inhomogeneous and low H concentration, No. 3 exhibited clear zero resistance at 44.5 K. To confirm its bulk superconductivity state, we measured magnetization as a function of $T$. Fig. 6(c) shows the results of FC and ZFC curves for No. 2 (blue) and No. 3 (red). From these results, we concluded that these sharp transitions in $\rho-T$ curves originate from a bulk superconducting state rather than surface superconductivity. This result is the first demonstration of superconductivity in Sm1111 epitaxial thin films achieved by hydrogen doping. The critical temperature $T_c$ determined from electronic and magnetic measurements is summarized in Fig. 6(d) and Table I. $T_c^{onset}$ of our films fit well to the dome-shaped superconducting phase obtained from polycrystalline bulks, even though the absolute values of $T_c$ were slightly lower compared to those of the bulk samples probably due to the influence of strain introduced via thin-film growth process. The shielding volume fraction (SVF) was small for No. 2 (1 %), whereas that of No. 3 (24.5 %) was enough to judge its bulk superconductivity although it was smaller compared to the SVF in polycrystalline bulks (79 % for $x = 0.13$ and 77 % for $x = 0.34$) [11]. This relatively small SVF could be attributed to the inhomogeneous H concentration along the film thickness as observed by SIMS [Fig. 4(c)].

Finally, we investigated the superconducting properties of the H-doped Sm1111 films (Nos. 2 and 3) under magnetic fields. Figures 7(a) and 7(b) show $\rho-T$ curves of No. 2 and No. 3 under external magnetic fields at $\mu_0 H = 0-9$ T applied along the $c$ and $a$ axes of the epitaxial films, respectively. The superconductivity in Nos. 2 and 3 maintained under 9 T both for in-plane and out-of-plane. A rough estimation of the orbital upper critical field, $\mu_0 H_{c2}(0)$, from Fig. 7(c) (sample 3) based on the Werthamer–Helfand–Hohenberg (WHH) theory [34] indicates 172 and 71 T along $a$- and $c$-axes, respectively. The corresponding coherence lengths in the $ab$ plane ($\xi^{//a}$) and along the $c$ axis ($\xi^{//c}$) would be 1.4 and 2.2 nm, respectively. We point out here that these values, which rely on the slopes ($\mu_0 dH_{c2}/dT$) at $T_c$, are only first estimations. In multiband superconductors (like it is expected to be the case for H-doped Sm1111) the WHH relation between the slope ($\mu_0 dH_{c2}/dT$) at $T_c$ and $\mu_0 H_{c2}(0)$ is necessarily not valid. For a more realistic description of the upper critical field high magnetic field data beyond 9T is necessary. The anisotropy factor ($\gamma$) estimated from $H_{c2}^{//a}$ and $H_{c2}^{//c}$ near $T_c$ was 2.4 for No. 3, which is comparable with that obtained in F-doped Sm1111 epitaxial films [23] but larger than 1.4 for 122-type K-doped BaFe$_2$As$_2$ [35]), owing to insulating SmO blocking layers rather than Ba lattices between the Fe$_2$As$_2$ layers. Due to the non-negligible H-concentration gradient, the estimated anisotropy factor in sample 3 may deviate slightly from the intrinsic value of $\gamma$ for SmO$_{1-x}$H$_x$FeAs with $x = 0.35$.

IV. CONCLUSION

We demonstrated high-$T_c$ superconductivity up to 48 K of H-doped 1111-type SmFeAsO epitaxial films on MgO single-crystalline substrates, which were grown by a topotactic chemical reaction between undoped epitaxial SmFeAsO precursor films deposited by Nd:YAG PLD and CaH$_2$ polycrystalline powders using post-deposition thermal annealing. The completely disordered H-doping at the O sites was experimentally found by ABF-STEM microstructure observation. The maximum H concentration with respect to O was ~35 %. Since this technique is very easy, it has the potential to become an effective and general method in the fabrication of various high-quality oxyhydride epitaxial films. Further optimization is needed in order to better control and homogenize the hydrogen content. The relationship between the onset $T_c$ and the maximum H concentration in the films was consistent with the previously established electronic phase diagram of H-doped SmFeAsO polycrystalline samples. Magnetization measurements validate that the observed superconductivity originates from the bulk state in the films, rather than from a surface state. The successful fabrication of H-doped Sm1111 films enables future investigations of their electronic structure by angle-resolved photoemission spectroscopy in order to elucidate the still controversially debated mechanism of superconductivity in Fe-based superconductors. Currently, investigations of the critical current densities of the high-$T_c$ 1111-type epitaxial films are carried out in order to determine important application-related properties as high-field magnets.

**ACKNOWLEDGMENTS**

This work was supported by the Ministry of Education, Culture, Sports, Science, and Technology (MEXT) through the Element Strategy Initiative to Form Core Research Center. The authors thank Soshi Iimura for providing us a standard polycrystalline bulk sample, SmFeAs($O_{0.4}H_{0.6}$), to perform quantitative chemical composition analysis by SIMS. T. K. was also supported by PRESTO, Japan Science and Technology Agency (Grant No. JPMJPR16R1). H. Hi. was also supported by the Japan Society for the Promotion of Science (JSPS) through Grants-in-Aid for Scientific Research (A) and (B) (Grant Nos. 17H01318 and 18H01700), and Support for Tokyotech Advanced Research (STAR).

desorption spectroscopy profile.

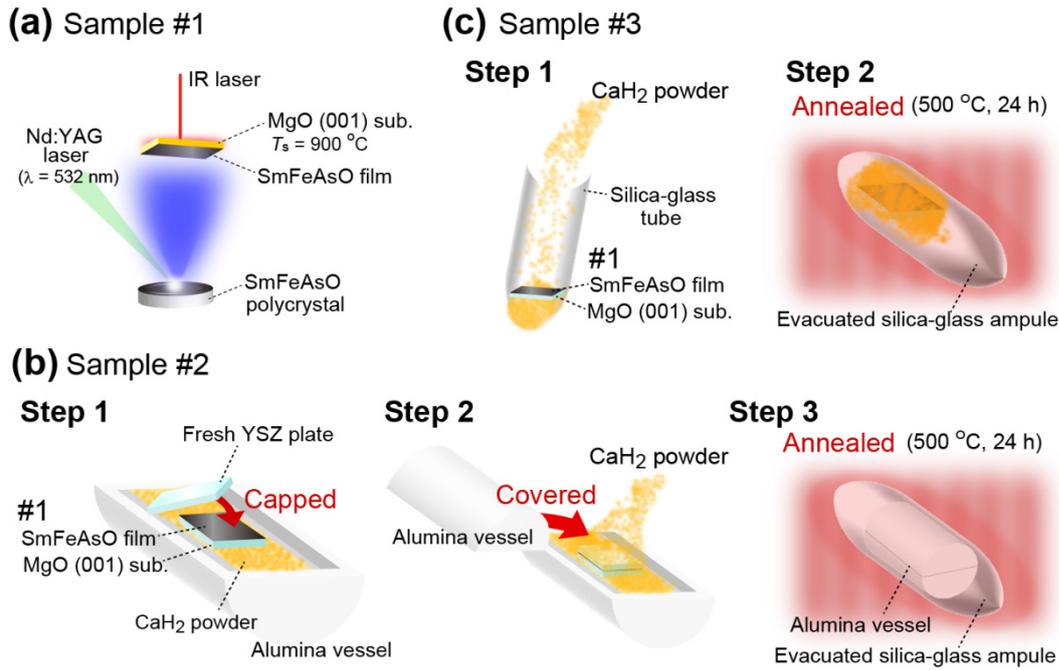

FIG. 1. Fabrication procedure of H-doped SmFeAsO (Sm1111) epitaxial thin films. (a) A schematic image for growth of an undoped Sm1111 epitaxial precursor film (No. 1) by Nd:YAG PLD under high vacuum. (b) H doping of the Sm1111 epitaxial precursor film. In step 1, the precursor (No. 1) was capped by a YSZ single-crystalline plate and put into an alumina vessel filled with $CaH_2$ powder. In step 2, the alumina vessel is closed. In step 3 the vessel was subsequently sealed in a silica-glass ampule and thermally annealed in vacuum at 500°C for 24h. The resulting H-doped Sm1111 film is denoted as No. 2. (c) Heavily H-doped Sm1111 epitaxial thin film. In step 1, the precursor film was inserted into a silica-glass tube that was filled with $CaH_2$ powder. In step 2, the tube was sealed under vacuum and then thermally annealed at 500 ºC for 24h. The heavily H-doped Sm1111 film is denoted as No. 3.

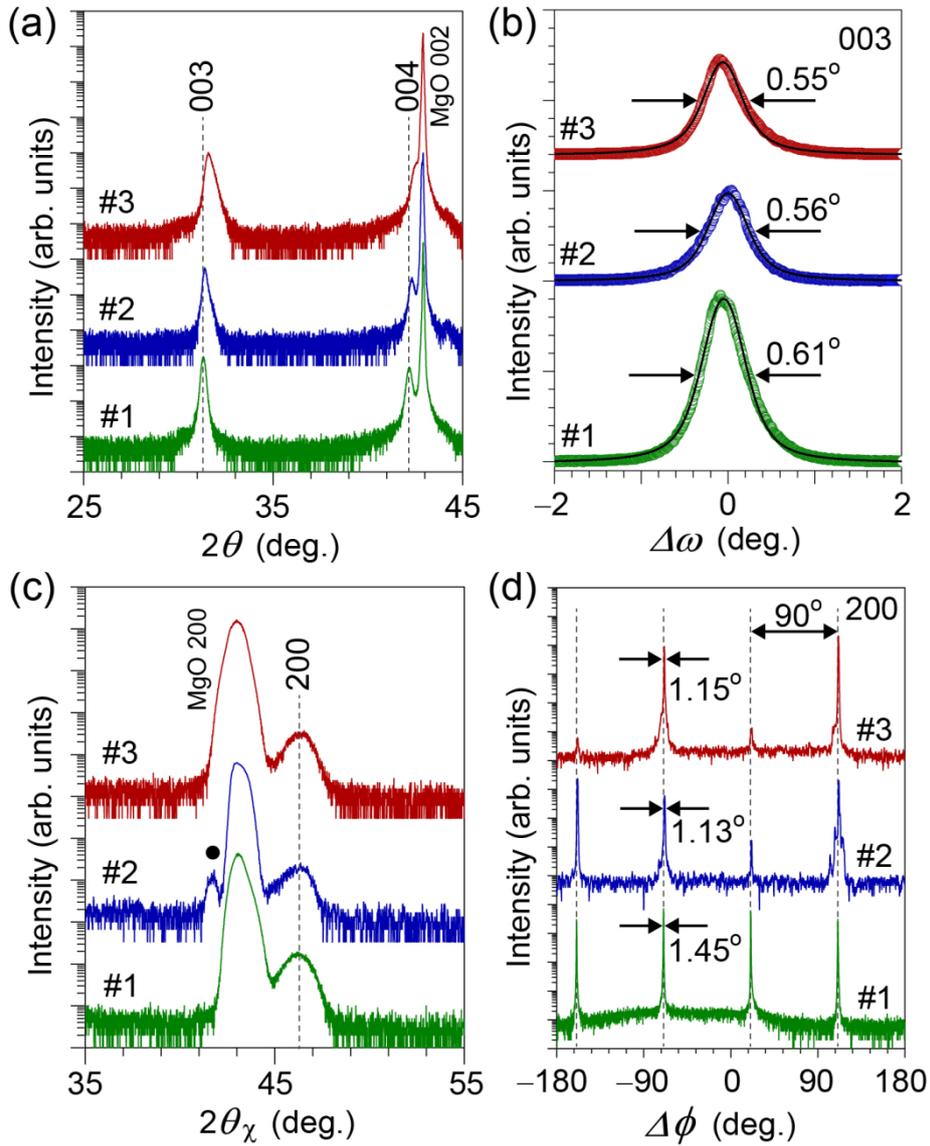

FIG. 2. Structural analyses of the undoped precursor film (No. 1) and H-doped Sm1111 epitaxial thin films (Nos. 2 and 3). (a) out-of-plane XRD patterns (b) 003 rocking curves of samples 1 – 3. (c) in-plane XRD patterns of samples 1 – 3. Filled circle in (c) represents a diffraction of an impurity phase such as $As_2O_5$ 400. (d) In-plane symmetry of Sm1111 200 diffractions.

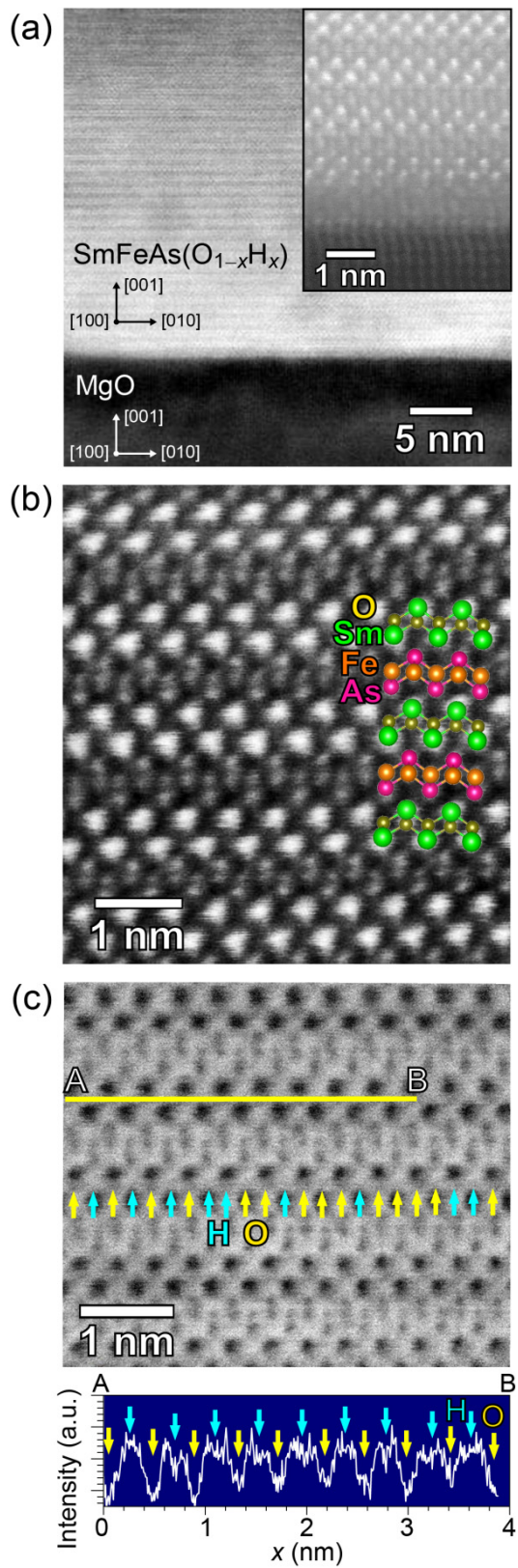

FIG. 3. Cross-sectional microstructure analysis of the sample 3. The incident electron

beam was along the *a* axis of Sm1111. (a) Wide-view HAADF-STEM image. The inset shows atomic resolution interface between the film and MgO. (b) The atomic-resolution HAADF-STEM image in the Sm1111 film region. (c) atomic-resolution ABF-STEM image in the Sm1111 film region. O and H atoms at O sites are indicated by yellow and light blue arrows, respectively. The bottom panel is the intensity plot along the line A – B in the upper ABF-STEM image.

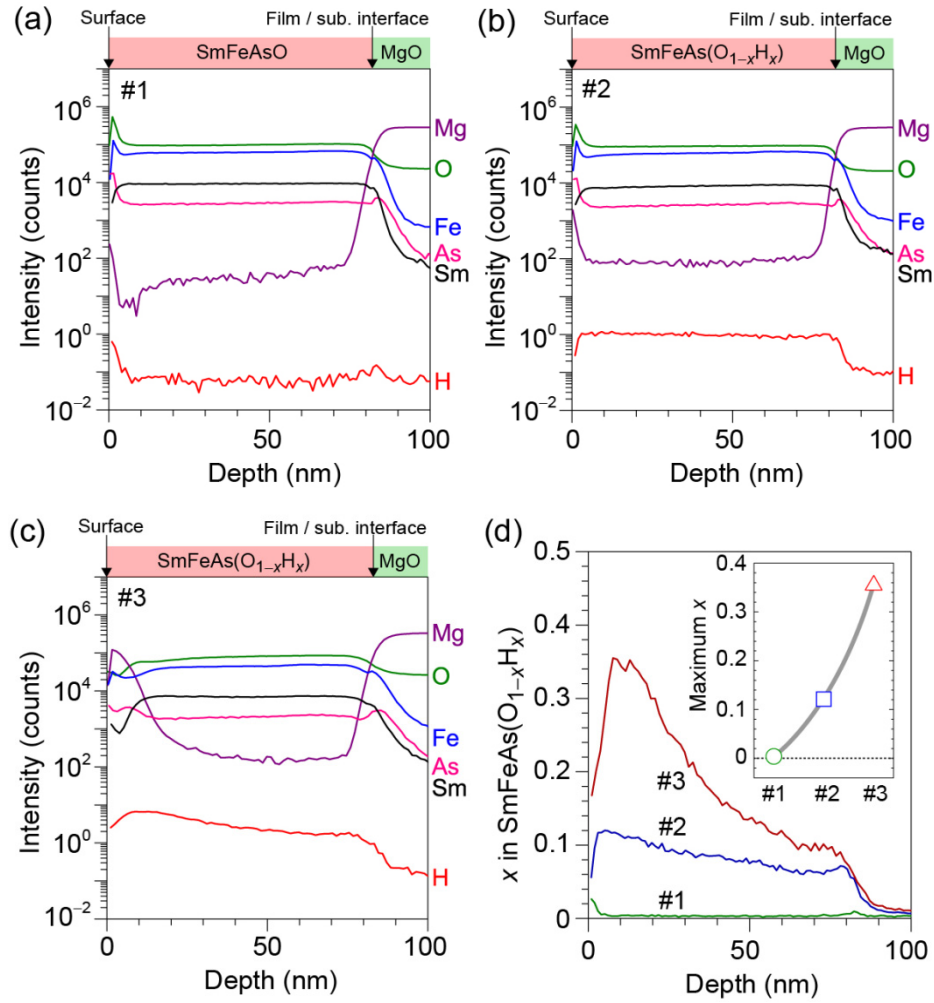

FIG. 4. H concentration ($x$) in SmFeAs($O_{1-x}H_x$) epitaxial thin films. SIMS profiles of (a) No. 1, (b) No. 2, and (c) No. 3 as a function of depth from surface, i.e., depth of 0 nm denotes the top of the film surface. These profiles were normalized with respect to the Fe signal. (d) Depth dependence of $x$ for all three films (Nos. 1 – 3). The inset shows the maximum $x$ for each sample.

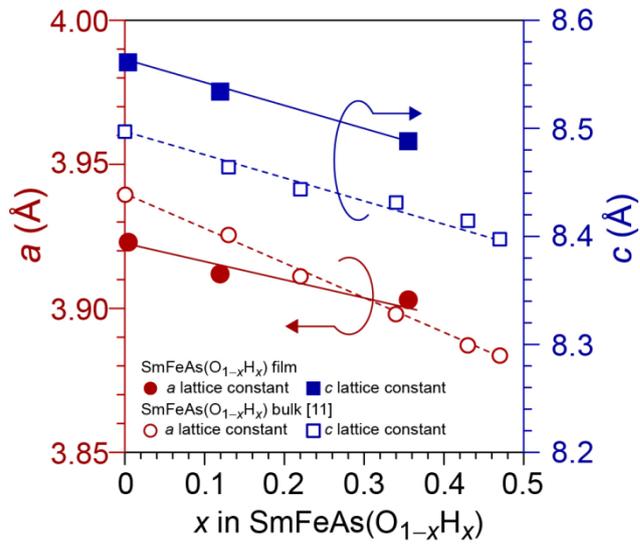

FIG. 5. Relationship between *x* and lattice parameters. Filled red circles and blue squares denote *a*- and *c*-axis lattice parameters of fabricated H-doped Sm1111 epitaxial films, where H contents around film surface were employed as *x*. Open red circles and blue squares correspond to those of H-doped Sm1111 polycrystalline bulks [11].

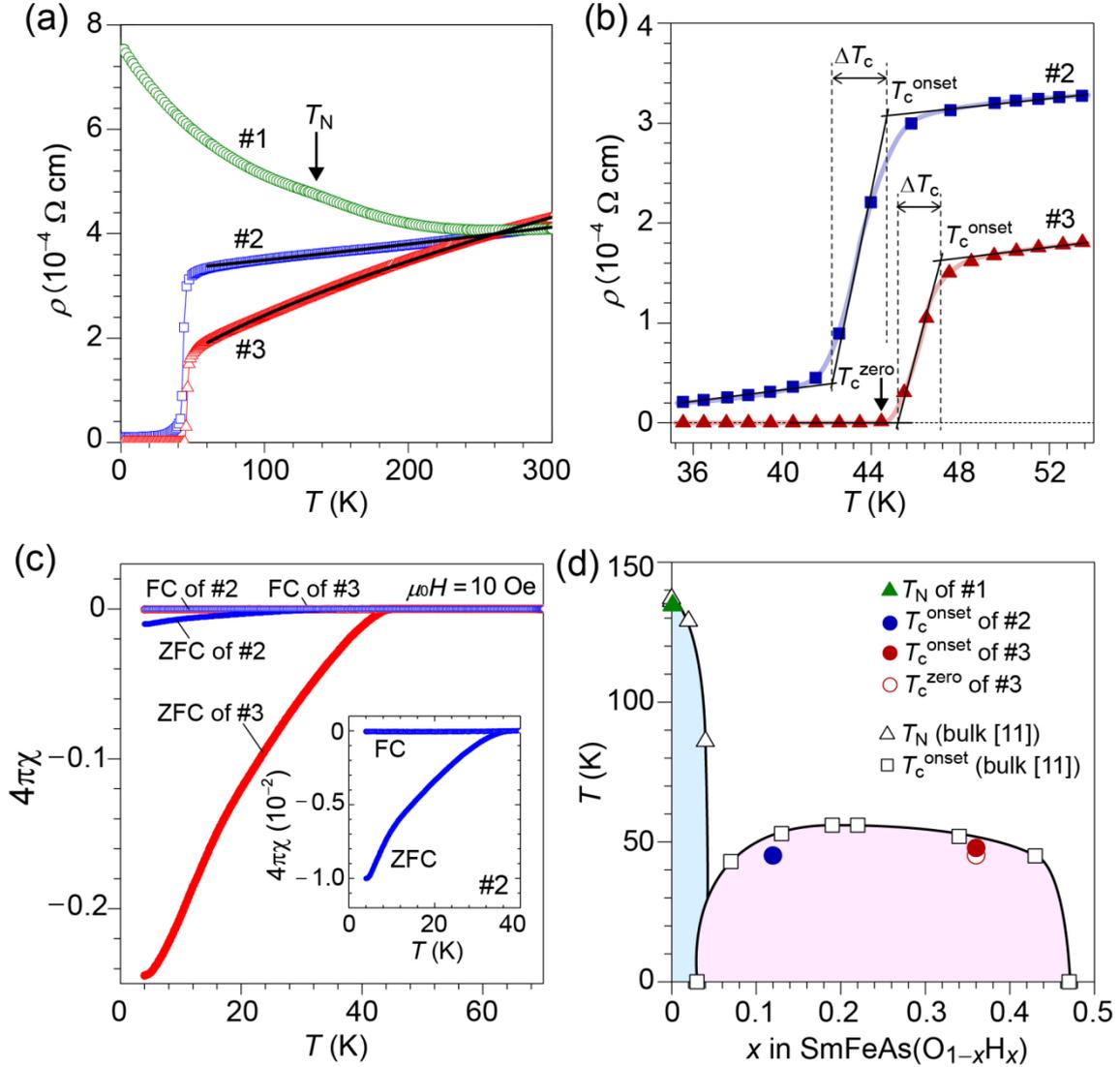

FIG. 6. Superconductivity of epitaxial H-doped Sm1111 thin films. (a) Temperature ($T$) dependence of electrical resistivity ($\rho$) of Nos. 1 – 3 from 2 to 300 K. Néel temperature ($T_N$) for No. 1 is the $\rho$ anomaly temperature due to antiferromagnetic transition. Black lines for Nos. 2 and 3 represent fitting results using a power-law equation, $\rho = \alpha + \beta T^n$, where $\alpha$ and $\beta$ are constants. (b) An enlarged plot of (a) around $T_c$. $T_c^{onset}$ and $T_c^{zero}$ were defined as $T$ intersection of two fitted linear lines as shown in (b). The transition width $\Delta T_c$ is defined as difference between $T_c^{onset}$ and $T_c^{zero}$. (c) $T$ dependence of the magnetization of Nos. 2 and 3 through ZFC and FC measurements under an applied magnetic field of 10 Oe. The inset shows an enlarged view for No. 2. (d) The electronic phase diagram of H-doped Sm1111 as a function of $x$. Open squares and triangles

represent $T_c$ and $T_N$ of H-doped Sm1111 bulks that are reported in ref [11]. The filled triangle indicates $T_N$ of No. 1. Filled circles denote $T_c^{onset}$ of H-doped Sm1111 epitaxial thin films, with $x$ taken as the maximum H content observed near the film surface.

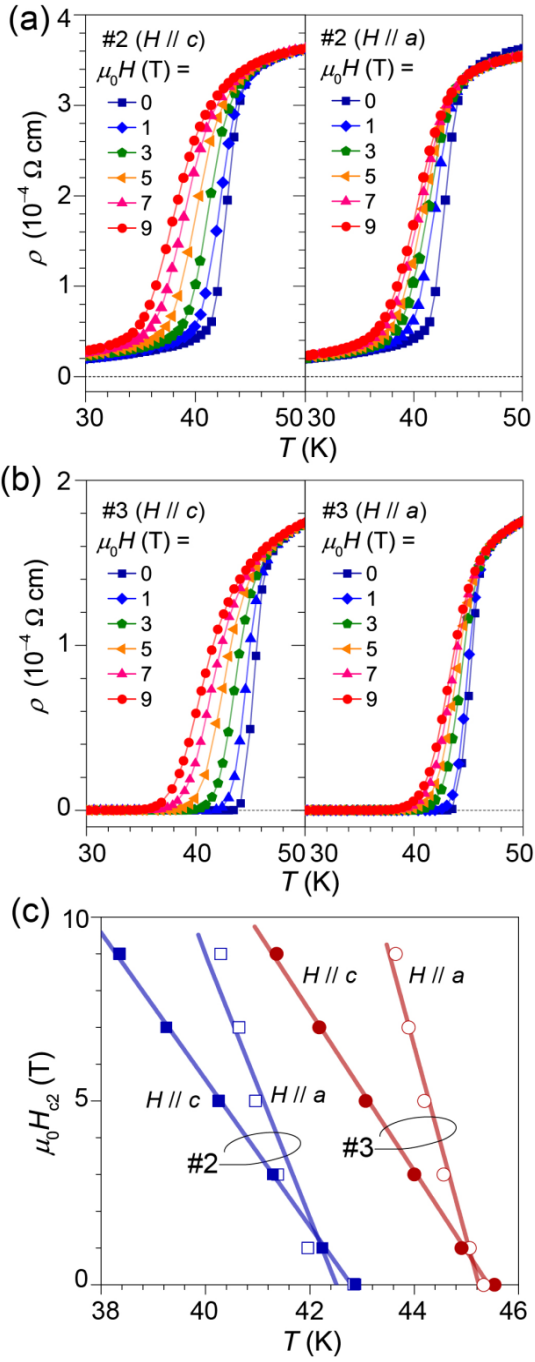

FIG. 7. $\rho-T$ curves under magnetic field ($\mu_0 H$) up to 9 T. (a) and (b) are results for Nos. 2 and 3, respectively. The applied magnetic fields were along $c$ axis for left panels and along $a$ axis for right panels. (d) Relationship between applied $\mu_0 H$ and $T_c^{50}$, which is defined as temperature where $\rho$ becomes 50 % of the resistivity of the normal state above $T_c$.

TABLE I. Summary of superconducting properties of the H-doped Sm-1111 epitaxial films. Anisotropy factor $\gamma$ is $\gamma = (m_c / m_{ab})^{1/2} = \xi_{ab} / \xi_c = H_{c2}^{//ab} / H_{c2}^{//c}$, where $m$ and $\xi$ are effective mass and coherent length, respectively.

| Sample | $T_c^{onset}$ (K) | $T_c^{zero}$ (K) | $\Delta T_c$ (K) | $T_c^{mag}$ (K) | SVF (%) | $n$ |
|---|---|---|---|---|---|---|
| **No. 2** | 45.2 | – | 3.4 | 37.1 | 1.0 | 1.1 |
| **No. 3** | 47.9 | 44.5 | 3.0 | 45.0 | 24.5 | 0.6 |

| | $H_{c2(0)}^{//c}$ (T) | $H_{c2(0)}^{//a}$ (T) | $\gamma$ | $\xi^{//c}$ (nm) | $\xi^{//a}$ (nm) |
|---|---|---|---|---|---|
| **No. 2** | 59 | 106 | 1.8 | 2.4 | 1.8 |
| **No. 3** | 71 | 172 | 2.4 | 2.2 | 1.4 |

Supplemental Material for "Superconductivity at 48 K of heavily hydrogen-doped SmFeAsO epitaxial films grown by topotactic chemical reaction using CaH$_2$"

Jumpei Matsumoto (1), Kota Hanzawa (1), Masato Sasase (2), Silvia Haindl (3), Takayoshi Katase (1), Hidenori Hiramatsu (1,2,a), and Hideo Hosono (2)

(1) Laboratory for Materials and Structures, Institute of Innovative Research, Tokyo Institute of Technology, Mailbox R3-3, 4259 Nagatsuta-cho, Midori-ku, Yokohama, Kanagawa 226-8503, Japan

(2) Materials Research Center for Element Strategy, Tokyo Institute of Technology, Mailbox SE-1, 4259 Nagatsuta-cho, Midori-ku, Yokohama, Kanagawa 226-8503, Japan

(3) Tokyo Tech World Research Hub Initiative (WRHI), Institute of Innovative Research, Tokyo Institute of Technology, 4259 Nagatsuta-cho, Midori-ku, Yokohama, Kanagawa 226-8503, Japan

(a) h-hirama@mces.titech.ac.jp

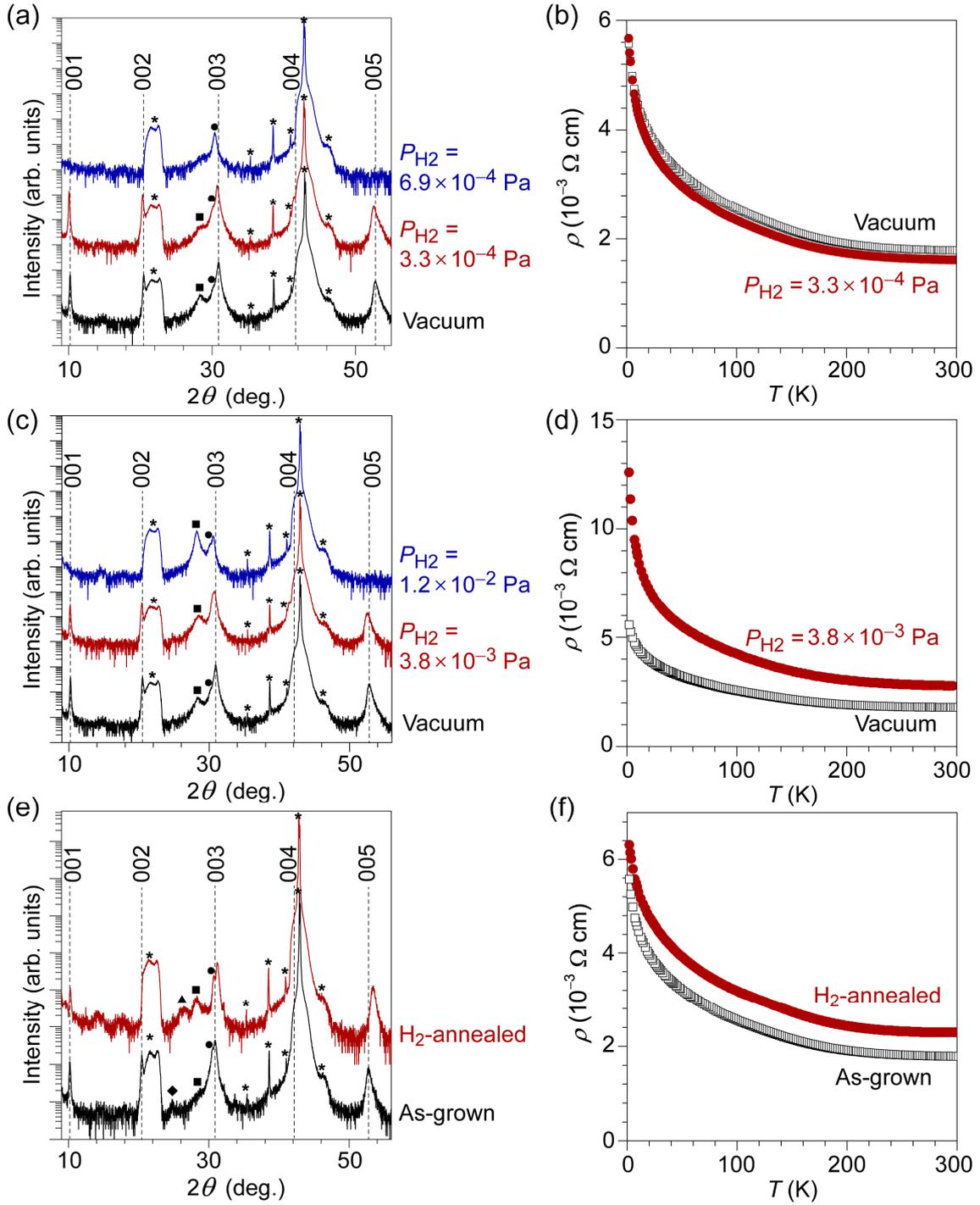

FIG. S1. Summary of preliminary experimental H-doping attempts for Sm1111 epitaxial films. All XRD patterns were observed by Bragg-Brentano geometry to identify small amounts of impurity phases. (a) XRD patterns for the as-grown Sm1111 film deposited under high vacuum by Nd:YAG PLD (black curve) and films deposited under different partial hydrogen gas pressure ($P_{H2}$) (red and blue curves) during PLD growth. (b) Temperature ($T$) dependence of electrical resistivity ($\rho$) of the obtained films

in (a). (c) XRD patterns of Sm1111 films grown by Nd:YAG PLD at different $P_H$, where H radical gas was introduced through an rf-plasma generator operating at 150 W. (d) $\rho-T$ curves of samples shown in (c). (e) Effect of post-deposition thermal annealing at 800 ºC under $H_2$ atmosphere for as-grown Sm1111 epitaxial films. The internal $H_2$ gas pressure in the furnace was maintained to be 1 – 2 MPa. (f) $\rho-T$ curves for as-grown and H-annealed Sm1111 in (e). In (a), (c), and (e), dotted lines, astarisks, circles, squares, diamond, and triangle represent diffraction peak positions originating from Sm1111 00$l$, MgO substrate, SmAs 200, a possible impurity phase of $Sm_2O_3$ 222 or $Sm_3Fe_5O_{12}$ 400, $SmAsO_4$ 200, and $Sm_3Fe_5O_{12}$ 321, respectively.

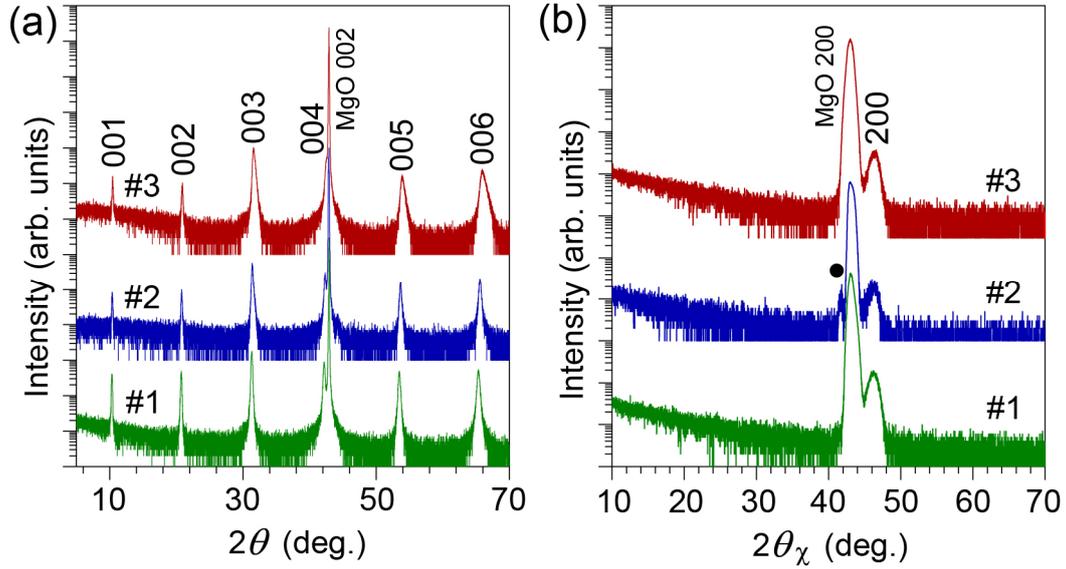

FIG. S2. Wide-angle range XRD patterns of Figs. 2a for (a) out-of-plane and 2b for (b) in-plane. Filled circle indicates a diffraction attributed to an impurity phase, possibly $As_2O_5$.

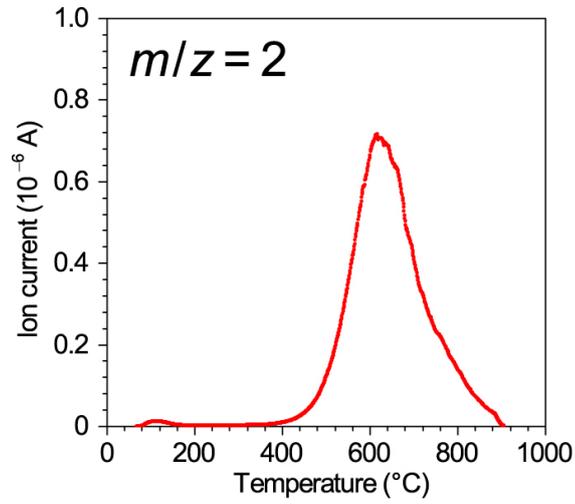

FIG. S3. Thermal desorption spectroscopy profile of hydrogen molecule ($m/z = 2$) from CaH$_2$ powder.

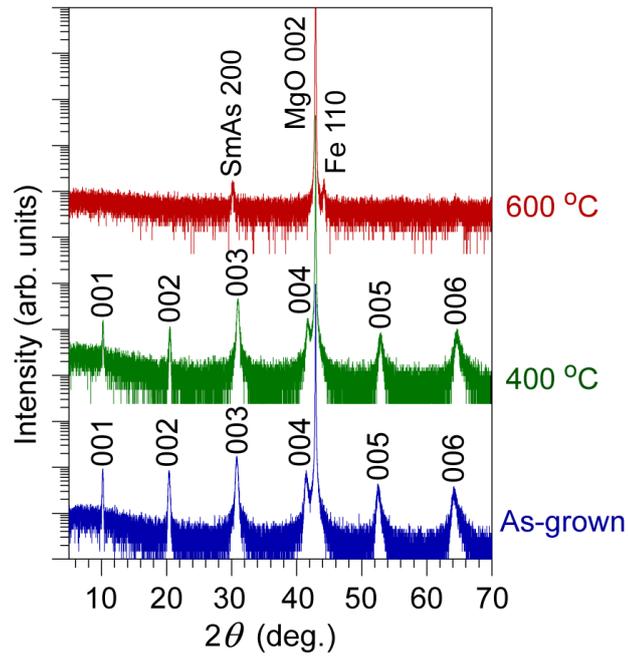

FIG. S4. XRD patterns of thermally annealed films at 400 ºC and 600 ºC. The experimental setup is the same as that for sample #3. That of an as-grown film (i.e., before annealing treatment) is shown for comparison.

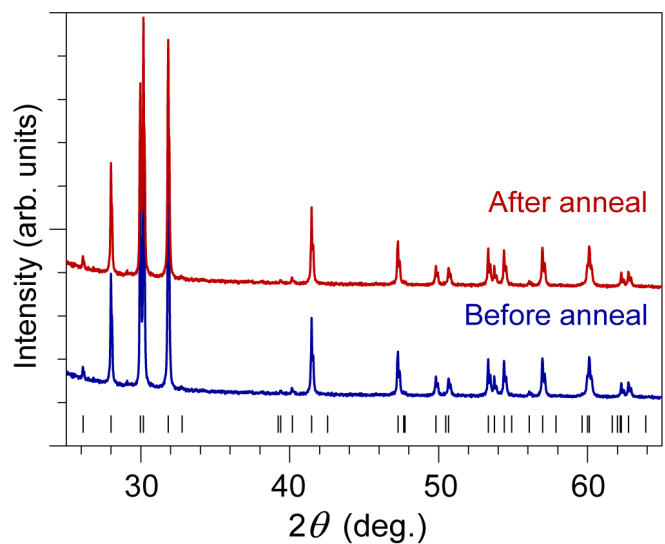

FIG. S5. XRD patterns of CaH$_2$ powders before and after thermal annealing using the process for sample #3. Simulated peak positions are shown at the bottom by vertical bars.